\documentclass[a4paper,12pt]{article}
\usepackage{a4wide}
\usepackage[bookmarks=false,colorlinks]{hyperref}
\usepackage[pdftex]{graphicx}
\usepackage{amsmath,amsfonts,amssymb,mathtools,bm}
\usepackage{rotating,amsmath,bm}
\usepackage{tabularx}
\usepackage[table, dvipsnames]{xcolor}
\usepackage{multicol,multirow}
\usepackage{latexsym}
\usepackage{relsize}
\usepackage[center]{subfigure}
\usepackage{float}
\usepackage{caption}
\usepackage{slashed}
\usepackage{lineno}
\usepackage{cite,mcite}
\usepackage{appendix}
\usepackage{url}
\usepackage{placeins}
\hypersetup{urlcolor=blue, citecolor=blue}
\usepackage[english]{babel}
\usepackage{bbold}
\usepackage{wasysym}
\usepackage[normalem]{ulem}
\usepackage{tikz}
\usetikzlibrary{arrows}
\tikzstyle{int}=[draw, minimum size=3em]
\tikzstyle{init}=[pin edge={to-, thick, black}]

\graphicspath{{data/}, {data_2/}, {data_3/}, {data_idq/}, {data_idq/15 mus/}}
\DeclareGraphicsExtensions{.pdf,.png,.jpg}

\date{\today}
 
\normalsize

\begin{document}

\title{\bf Investigating the coherent state detection probability of InGaAs/InP SPAD-based single-photon detectors}

\maketitle

{\centering \sc
Andrey Koziy$^{1}$, Andrey Tayduganov$^{1,2}$, Anton Losev$^{1,2,3}$, Vladimir Zavodilenko$^{1,2}$, Alexander Gorbatsevich$^{3,4}$ and  Yury Kurochkin$^{1,2,5}$
\par}

\vspace{5mm}
{\centering \sl\small \noindent
$^1$~"Qrate" LLC, St. Novaya, d. 100, Moscow region, Odintsovo, Skolkovo village, 143026, Russia. \\
$^2$~NTI Center for Quantum Communications, National University of Science and Technology MISiS, Leninsky prospekt 4, Moscow, 119049, Russia\\
${}^3$~National Research University of Electronic Technology MIET, Shokin Square, 1, Zelenograd, 124498, Russia. \\
${}^4$~P.N. Lebedev Physical Institute of the Russian Academy of Sciences,  Leninsky prospect, 53, Moscow, 119333, Russia.\\
$^5$~Russian Quantum Center, Skolkovo, Moscow 143025, Russia
\par}

\vspace{1cm}

\begin{abstract}
In this work we investigate the probabilities of detecting single- and multi-photon coherent states on InGaAs/InP sine-gated and free-run single-photon avalanche diodes. As a result, we conclude that multi-photon state detection cannot be regarded as independent events of absorption of individual single-photon states. However, if a greater number of states are one- and two-photon, then an Independent model is permissible. We conclude that physical processes must occur in the diode structure, determining the correlation when several photons are absorbed simultaneously. We present two models that we can use to describe photon interactions in a structure. The Dependent model is based on an increase/decrease in the probability of detecting an $n$\textsuperscript{th} photon upon an unfortunate detection of an ($n-1$)\textsuperscript{th} photon in an $n$--photon state and is well physically grounded; The Empirical model offers a simple and accurate empirical relationship.
\end{abstract}

\thispagestyle{empty}

\newpage
\setcounter{page}{1}


\section{Introduction}

\qquad Single-photon detectors (SPD) every year more and more enter our lives in entirely different areas. These devices have found tremendous success in quantum key distribution (QKD) -- it is irreplaceable equipment in this field \cite{kiktenko2017demonstration, fan2020optimizing, zhang2020long}. There is an application to the time resolved emission measurements (TRE), where the SPD can be used to: check individual circuit elements' performance without directly affecting them \cite{stellari2013superconducting}, for singlet-oxygen luminescence detection \cite{boso2016time}, and other interesting applications \cite{zhang2019potential}. There are other areas of technology, for example, quantum computing on photons \cite{you2020superconducting}, the LIDAR system \cite{beer2018background}, fluorescence microscopy \cite{slenders2021confocal}, and many others. However, in these areas, SPD's use allows us to increase the measurement accuracy exclusively but is no a cornerstone in these systems' work.

In this paper, we consider InGaAs/InP single-photon avalanche diode (SPAD) based SPDs \cite{sanzaro2016ingaas}, the main application of QKD use. A big problem in SPD is the roughness in determining their key parameters -- this is metrological difficulties. However, we use these devices in an information security system designed to ensure secrecy due to nature's laws. Significant errors in determining the detection probability lead to the fact that the entire QKD system's parameters become more difficult to predict and acquire significant errors \cite{zhao2021practical}. For example, an incorrectly determined detection probability can lead to the fact that the system's absolute security will be an order of magnitude lower than expected. Otherwise, we can define the same detection probability too pessimistically. We can design the system with excessive security, which will negatively affect generating the secret key.

This paper considers the SPD parameter, called the photon detection probability (PDE). It is the probability that a single photon impinges the detector will be registered. Obtaining true single-photon sources is not a trivial task, and now these devices exist only as laboratory samples \cite{ripka2018room, wang2018bright}. For this reason, we currently use laser radiation sources in QKD systems \cite{min2020laser}. A significantly weakened photon beam generated by a laser has the properties of so-called quantum light. Poisson's law is valid for the probability of the existence of $k$ photons in a pulse from the energy of a given pulse $\mu$ \cite{liao2020direct}. We write this statement as follows:

\begin{equation}
	P(\mu, k) = \frac{\mu^k}{k!} e^{- \mu},
\end{equation}

with $P$ -- the probability of realizing the $k$ photons in the laser pulse.

In a widely used model for assessing a QKD system's security, we assume that $k$ photons in one laser pulse are detected independently of each other. From this assumption, we derive the following equation for the probability of laser pulse detection $P_{det}$ \cite{jin2019receiver}:

\begin{equation}
	P_{det} = 1 - e^{- \mu \eta},
\end{equation}

with $\eta$ -- the probability of detecting the single photon.

In fact, with the simultaneous realization of two, three, or more photons in the SPAD absorption region, various physical processes can occur, which will reduce or, conversely, increase the probability of detection.

We investigate the possibility of applicability of the model of independence of incoming photons. We put forward models that allow taking this phenomenon into account both from physical processes and an empirical approach. The studied SPD's had both gated and freerun modes and were tuned to have a minimum dark count rate (DCR) ($< 1$ kHz), with different afterpulse probabilities ($0.6 \ \%$, $3.2 \ \%$ and $11.7 \ \%$), with a dead time of about ($5 - 20 \  \mu\text{s}$).

Also we present the theoretical equations for accurate calculation of laser pulse detection probability through the counting statistics, that include the dead time, DCR and afterpulsing behaviour. In the work \cite{lopez2020study} authors presented the quite good analythical approach for calculating this probability, but they didn't take the afterpulsing effect into account. On the countrary, in our work we pay special attention to the afterpulse effect and use two models for it's accurate  accounting. In the work \cite{sarbazi2018impact} authors describe the SPAD processes by Markov chain. The main disadvantage is that in such model we need to calculate the evolution of the SPAD system to it's steady state, that is not-trivial.  In the work \cite{straka2020counting} authors authors describe the SPAD processes by Poisson point processes. The main disadvantage is similar as for Markov chain approach -- authors used the stochastic approach to simulate the evalution detector performance. On the countrary, in our work we use the analythical equations to analyzing the experimental counting statistics, and we don't use the difficult numerical calculations.

\section{Metrology approach and experimental data processing algorithm}

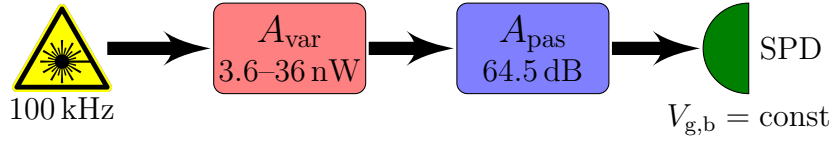
\begin{figure}[t!]\centering
\begin{tikzpicture}[scale=4, >=latex']
	\node[inner sep=0pt] (laser) at (0, 0){\includegraphics[width=13mm]{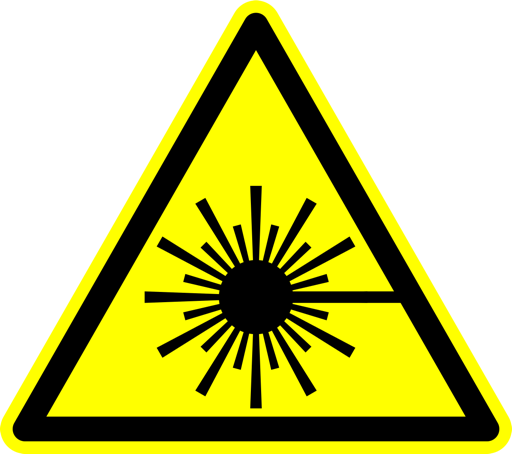}} node[below=5mm]{100\,kHz};
	\filldraw[fill=red!50, rounded corners] (0.5,-0.15) rectangle +(0.5,0.3) node[left=10mm,below=0.2mm]{\large$A_{\rm var}$} node[left=10mm,below=6mm]{3.6--36\,nW};
	\filldraw[fill=blue!50, rounded corners] (1.3,-0.15) rectangle +(0.5,0.3) node[left=10mm,below=0.2mm]{\large$A_{\rm pas}$} node[left=10mm,below=6mm]{64.5\,dB};
    \filldraw[fill=green!50!black] (2.26,0.15) arc (90:270:1.5mm) node[above=6mm,right]{SPD} node[below]{$V_{\rm g,b}={\rm const}$} -- cycle;
    \draw[black,line width=5, ->] (0.15,0) -- (0.49,0);
    \draw[black,line width=5,->] (1.01,0) -- (1.29,0);
    \draw[black,line width=5,->] (1.81,0) -- (2.1,0);
\end{tikzpicture}
\caption{Schematic optical circuit.}
\label{fig:scheme_1}
\end{figure}

\qquad We have carried out the experiments on the installation, which is schematically shown in the figure  \ref{fig:scheme_1}. On this picture, laser pulses has repetition rate $\nu_l = 100$ kHz and $\text{FWHM} \approx 50$ ns. After that, laser pulses come on the attenuator with power control ($A_{var}$), where it controls the output integral power (for $1$ s). In presented system, to obtain  $\mu = 0.1$ ph/pulse  we need to set power to $W \approx 3.6$ nW, and to obtain $\mu = 1$ ph/pulse we need to set power to  $W \approx 36$ nW. This value determines by the attenuation of $A_{var}$ equals $64.5$ dB, which includes the attenuation of the second attenuator and losses in the contacts and optical fiber. During all measurements with changing the $W$, we did not change the detector's parameters $V_g$ and $V_b$ (gate voltage and bias voltage, that determine the detector's characteristics).


\newpage

In the experiment, we need to get the next data sets:

\begin{itemize}
	\item $R^\prime$ -- count rate with laser off;
	\item $R_i$ -- count rate with laser on and output power $\mu_i$ from the set $\mu_i \in \{\mu_1, \hdots \mu_N\}$. 
\end{itemize}

From these statistics, we need to extract the count rate of the photon triggers $R^0_{sig (i)}$ (exclude the DCR and afterpulses), which can be derived as the sum of the occurred photon triggers on the detector $R_{sig(i)}$ and that could be registered. However, the dead time blocked the detector sensitivity $R^\prime_{sig}$. Great difficulties arise when we want to take into account the afterpulse influence of the statistics. Based on the definition, afterpulse is the probability that the detector's internal processes after one trigger will lead to the next trigger \cite{wang2019afterpulsing}. Afterpulse can generate afterpulse, and so on -- this is a recurrent process. For the analytical approach, we need to simplify this model -- this is how the first and second-order afterpulse approximation model was obtained (see eq. \ref{eq:2} in appendinx \ref{ap:det}).  We present the detailed derivation of these equations in our another work [reference to AP-DT]. So, for the low afterpulse probability ($p_{ap} < 2 \ \%$) we can use 1-st order model, but if $p_{ap} > 2 \ \%$, we need to use second-order model. In the data processing algorithm, we used only the second-order afterpulse model due to we need to obtain the high accuracy. From this equation we obtain the probability $P_{det(i)} = P^0_{sig(i)} = \frac{R^0_{sig(i)}}{\nu_l}$ that laser pulse, if one comes, would be detected:

\begin{equation}\label{eq:pdet}
	\begin{split}
		&P_{det(i)} = \frac{R_{sig(i)}}{\nu_l - R_i \frac{\tau}{T} + R_{sig(i)}(\frac{\tau}{T} - [\frac{\tau}{T}])}, \\
		\text{with } &R_{sig(i)} = \nu_{\tau} - \nu_{\tau} \frac{1 - \frac{1 - p^2_{ap}}{2p_{ap}} \biggl(1 - \sqrt{1 - \frac{4 p_{ap} \frac{R_i}{\nu_{\tau}}}{1 + p_{ap}}}\biggr)}{1 - \frac{1 - p^2_{ap}}{2p_{ap}} \biggl(1 - \sqrt{1 - \frac{4 p_{ap} \frac{R^\prime}{\nu_{\tau}}}{1 + p_{ap}}}\biggr) (1 - \frac{R_i}{\nu_{\tau}} \frac{\tau}{T})},
	\end{split}
\end{equation}

where $\nu_{\tau} = \frac{1}{\tau}$ -- the limit count rate of the detector (determined by the dead time $\tau$), $T = \frac{1}{\nu_l}$ -- period of laser pulses.

\section{Analyzing the $k-$ photon detection probability models}
\qquad Assuming mutual independence of photons in a $k$--photon state, the well known theoretical prediction is expressed as (below we will denote this model as "Independent" model) \cite{jin2019receiver}:
\begin{equation}
	P_{det}(\mu) =\sum_{k=1}^\infty {\mu^k \over k!} e^{-\mu}\big[ 1 - (1 - \eta)^k \big] = 1 - e^{-\eta\mu} \,.
	\label{eq:R_sig_th_1}
\end{equation}

where the term $1 - (1 - \eta)^k$ denotes the probability that at least one photon will trigger the detector, assuming that simultaneous detecting of $k$ photons is considered as joint and independent events. The term $\frac{\mu^k}{k!} e^{- \mu}$ denotes the probability that the laser pulse with energy $\mu$ will have $k$ photons.  

Note that from the detection probability we can easily obtain the detection rate $R^{0(th)}_{sig}$ for the theoretical model:

\begin{equation}
	R^{0(th)}_{sig} = \nu_l P_{det}. 
\end{equation}

If the $k-$ photons interaction processes inside the detector are not independent (but still be joint), we can derive the laser pulse detection probability as:

\begin{equation}
	P_{det}(\mu) \simeq \sum_{k=1}^n {\mu^k \over k!} e^{-\mu} \, \eta_k \,,
	\label{eq:R_sig_th_2}
\end{equation}
with some unknown detection probabilities $\eta_k$ (in particular, $\eta_1=\eta$).  In the independent model, we can derive this $\eta_k$ as:

\begin{equation}
	\begin{split}
		\eta_1 & = \eta, \\
		\eta_2 & = 2 \eta - \eta^2, \\
		\eta_3 & = 3 \eta - 3 \eta^2 + \eta^3, \\
		\vdots \\
		\eta_k & = 1 - (1 - \eta)^k. 
	\end{split}
\end{equation}

To introduce the model with accounting the dependence of interaction of the photons (below we will denote this model as "Dependent"), we will enter the parameter $\rho_k$ that amplifies (or reduce) the probability of detecting photon after the unsuccessful photon detection: 

\begin{equation}
	p_{\overline{\gamma}_1,... \overline{\gamma}_{n-1}} (\gamma_k) = \rho_k \eta,	
\end{equation}

with $\gamma_k$ -- is the event with successful detecting of $k$-th photon in single laser pulse, and $\overline{\gamma_k}$ -- is unsuccessful. The possible physical nature of collective effects phenomenon will be presented in the next paper.

Define the probability of detecting the  $2$, $3$ and $n$ photons in case, when $\eta_1 = \eta \rho_1$, and $\rho_1 = 1$:

\begin{equation}
	\begin{split}
		\eta_1 & = \eta, \\ 
		\eta_2 &= p(\gamma_1 \cup \gamma_2) = 1 - p(\overline{\gamma_1 \cup \gamma_2}) = 1 - p(\overline{\gamma}_1 \cap \overline{\gamma}_2) 	= 1 - p(\overline{\gamma}_1)p_{\overline{\gamma}_1}(\overline{\gamma}_2) = \\
		&= 1 - (1 - p(\gamma_1))(1 - p_{\overline{\gamma}_1}(\gamma_2)) = 1 - (1 - \eta)(1 - \rho_2 \eta) = \\
		&= \eta (1 + \rho_2) - \rho_2\eta^2 \,, \\
		\eta_3 &= p(\gamma_1 \cup \gamma_2 \cup \gamma_3) = 1 - p(\overline{\gamma_1 \cup \gamma_2 \cup \gamma_3}) = 1 - p(\overline{\gamma}_1 \cap \overline{\gamma}_2 \cap \overline{\gamma}_3) = \\
		&= 1 - p(\overline{\gamma}_1) p_{\overline{\gamma}_1}(\overline{\gamma}_2 \cap \overline{\gamma}_3) = 1 - p(\overline{\gamma}_1) p_{\overline{\gamma}_1}(\overline{\gamma}_2) p_{\overline{\gamma}_1, \overline{\gamma}_2}(\overline{\gamma}_3) = \\
		&= 1 - (1 - p(\gamma_1))(1 - p_{\overline{\gamma}_1}(\gamma_2))(1 - p_{\overline{\gamma}_1, \overline{\gamma}_2}(\gamma_3)) = 1 - (1 - \eta)(1 - \rho_2 \eta)(1 - \rho_3 \eta) = \\
		&= \eta (1 + \rho_2 + \rho_3) - \eta^2(\rho_2 + \rho_2 \rho_3 + \rho_3) + \rho_2 \rho_3 \eta^3 \,, \\
		&\vdots \\
		\eta_k &= 1 - \prod^k_{i=1}(1 - \rho_i \eta) = \eta \sum^k_{i = 1} \rho_i - \eta^2 \sum^k_{i,j = 1; j > i} \rho_i \rho_j + \eta^3 \sum^k_{i,j,k = 1; k>j>i} \rho_i \rho_j \rho_k \cdots .
	\end{split}
\end{equation}

There are some constrictions in this approach. Parameter $\rho_k$ denotes the amplification or loss of the probability of detecting the single-photon after $k-1$ unsuccessful detections. It's obvious, that $\rho_k > 0$.  From the point of view of physics of avalanche generation processes, all $\rho_k$ must be $> 1$ or $< 1$. There is no upper bound for the $\rho_i$, but there is the upper bound for $\eta_k$: $\eta_k < 1$. 

\begin{figure}[t!]\centering
	\includegraphics[width=0.45\textwidth]{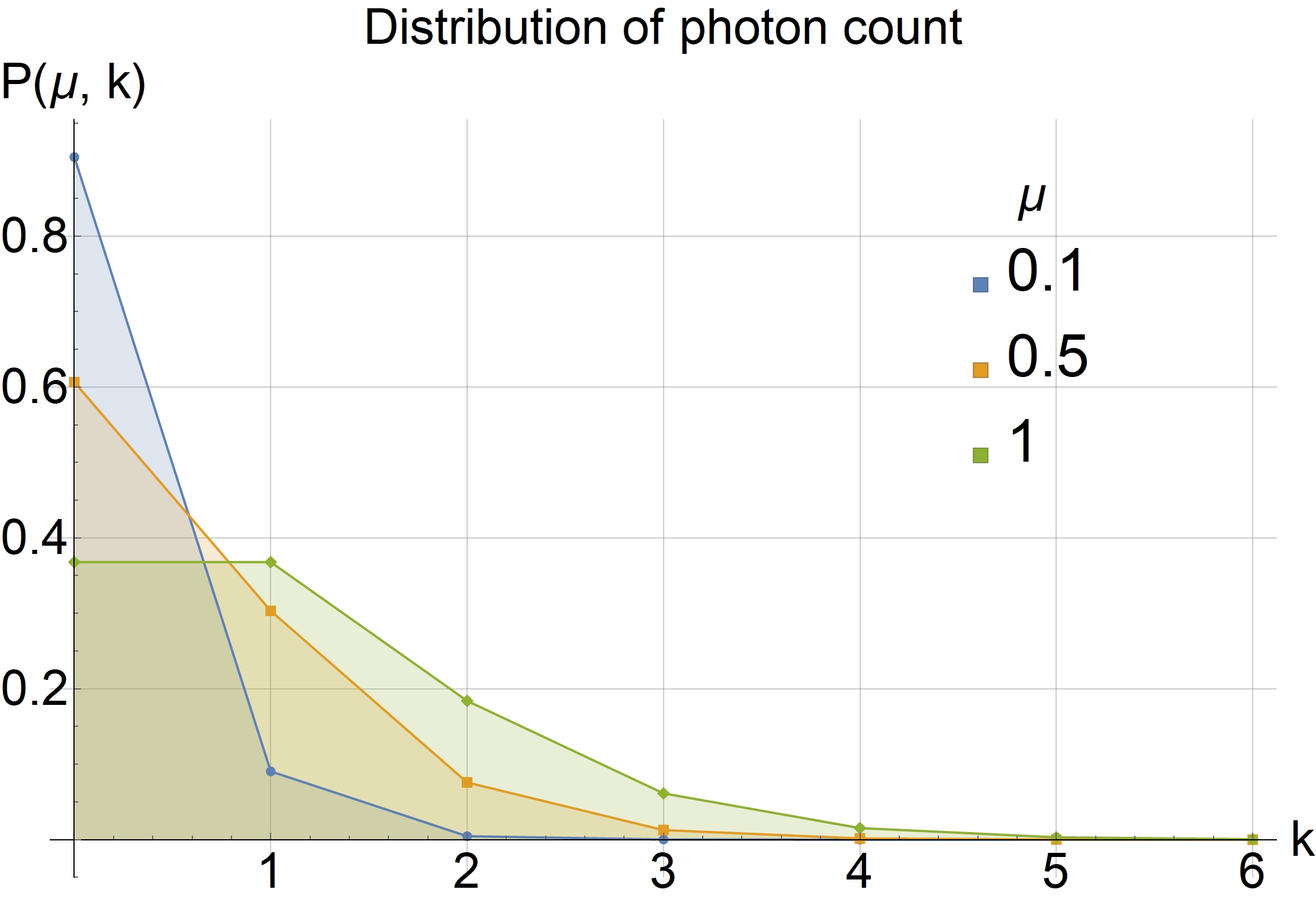} \hspace{3mm}
	\includegraphics[width=0.45\textwidth]{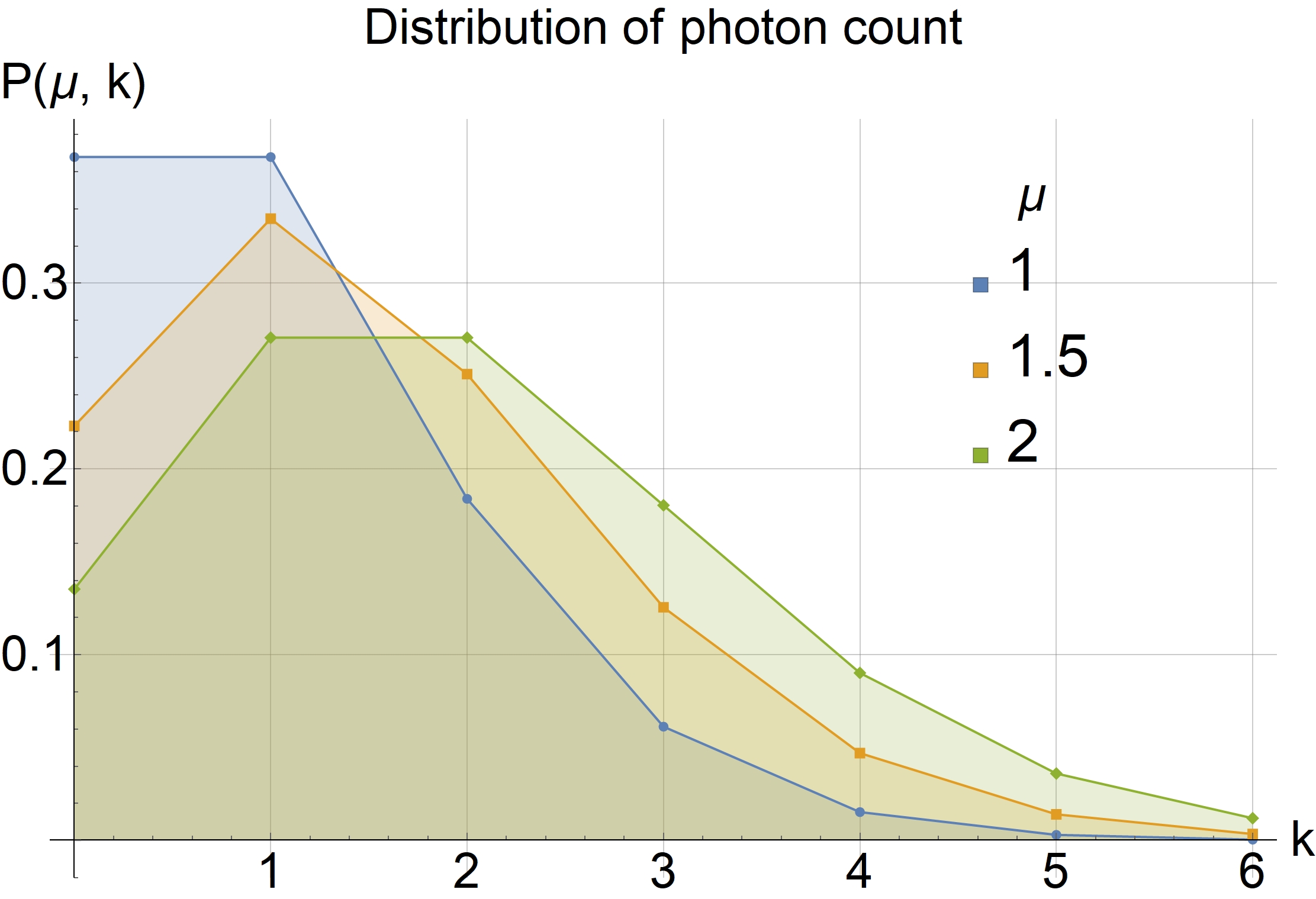}
	\caption{Probability distribution of photon count per laser pulse with different energies: a) $\mu = \{0.1, 0.5, 1 \}$; b) $\mu = \{1, 1.5, 2\}$. }
   \label{fig:poissons}
\end{figure}

The independent model has only one fitted parameter, which makes sense of PDE ($\{\eta \}$), and this model can be simply used in the theoretical models for QKD. The dependent model has a more robust physical description and more fitted parameters ($\{\eta, \rho_i\}$). The application for this model in theoretical studies is complicated, but the results should be more accurate. The main difficulties begin when the parameter $\mu$ is big enough (more than one ph/pulse). In this case, the number of used $\rho_i$ should be more than $3$. This statement can be displayed on the picture \ref{fig:poissons}. 

According to picture \ref{fig:poissons} we can conclude that for $\mu = 0.1$ ph/pulse, we have enough to consider only $\eta_1 = \eta$ term, because probability of realization of $2$ photons is negligible: $\mu = 0.1 \to \{\eta_1 \}$. If $\mu = 0.5 \to \{\eta_1, \eta_2\}$, if $\mu = 1 \to \{\eta_1, \eta_2, \eta_3\}$, if $\mu = 1.5 \to \{\eta_{1 \to 5} \}$, if $\mu = 2 \to \{\eta_{1 \to 6} \}$.  

If one model is too rough for great values for $\mu$, and another is too difficult, we will introduce the empirical model with only two fitted parameters: $\eta$ and $\rho$. We can write the next set of the $\eta_k$ parameters for the empirical model:

\begin{equation}
	\begin{split}
		\eta_1 &= \eta, \\
		\eta_2 &= 2 \eta - \rho \eta^2, \\
		\eta_3 &= 3 \eta - 3 \rho \eta^2 + \rho^2 \eta^3, \\
		\vdots \\
		\eta_k &= \frac{1 - (1 - \rho \eta)^k}{\rho}.
	\end{split}
\end{equation}

Moreover, for this model by analogy with an independent model, we can derive the analytical equation for $P_{det} (\mu)$:

\begin{equation}
	P_{det}(\mu) = \sum^\infty_{k = 1} \frac{\mu^k}{k!} e^{-\mu} \frac{1 - (1 - \rho \eta)^k}{\rho} = \frac{1 - e^{- \rho \eta \mu}}{\rho}. 
\end{equation}

Let us bring together the expression for deriving $\eta_k$ for three described models:

\begin{equation}
	\begin{split}
		\text{Independent: } &\eta_k = 1 - (1 - \eta)^k, \\
		\text{Dependent: } &\eta_k = 1 - \prod^k_{i = 1} (1 - \rho_i \eta), \\
		\text{Empirical: } &\eta_k = \frac{1 - (1 - \rho \eta)^k}{\rho}.
	\end{split}
\end{equation}

Now, we have three models that we will verify on the experimental data. As a criterion of the models' adequacy, we will use the $\chi^2$ function, presented below.

\section{Fit mathematics}

\qquad To make a decision, which model is more accurate for the experimental data, we introduce the universal criterion -- $\chi^2$ function, reduced to several degrees of freedom $r$:

\begin{equation}
	\chi^2_r = \frac{\chi^2}{r},
\end{equation}

where $r$  can be determined as the difference in the number of experimental points and number of fitted parameters. Parameter $\chi^2$ can be calculated as:

\begin{equation}
	\chi^2 = \sum^N_{i=1}\frac{(R^{0}_{sig(i)} - R^{th}_i)^2}{\sigma^2_{R^{0}_{sig(i)}} + \sigma^2_{p_i}},
\end{equation}

where $i$ -- is the current experimental point, $N$ -- total count of experimental points, $R^{0}_{sig(i)}$ -- mean value of the experimental data, $R^{th}_i$ -- theoretical value in the point, $\sigma_{R^{0}_{sig(i)}}$ -- standard deviation for the experimental data, $\sigma_{p_i}$ -- standard deviation due to uncertainties in laser pulse power.  A detailed calculation of these values is given in the appendix \ref{ap:chi2}. 

In the experiment, $\sigma_{R_i}$ is relatively small,  while the $\sigma_{p_i}$ value is one order of magnitude more. To improve the accuracy of the measurements, we need to ensure power stability and low uncertainties.   

We construct the fitting procedure in such a way as to minimize the $\chi^2$ function. As a criterion of the model's adequacy, we take the value of $\chi^2_r < 3$. If the $\chi^2_r \leq 1$ then the model approximates experimental data well. If $1 < \chi^2_r < 3$, then we think that the model can approximate data, but uncertainties will be significant enough. If $\chi^2_r > 3$, then we think that model is unusable for this experimental data.

Worth noting that we can use the one model that can range from $\mu$: $[0, 1]$ ph/pulse and unusable for $\mu > 1$ ph/pulse.

\section{Results}

\qquad Now we will analyze the results of measurements and data processing for the three detectors: two custom sinusoidal gated detectors SPD1 and SPD2 (with gated frequency $\nu = 312.5$ MHz), based on InGaAs/InP SPADs (manufactured by Wooriro company; SPAD1: PA19H262-0006 and SPD2: MF20D300-0001 with the butterfly housing and a built-in enclosure cooling system) in the gated mode based on the SPAD, operational temperature $T = -50 \ {}^\circ$C; and one freerun detector from ID Quantique (IDQ) also InGaAs/InP SPAD based, operational temperature $T = -90 \ {}^\circ$C. Investigating the detectors in gated and freerun modes will allow us to say whether the presented equations are universal or better suited for a specific detector type. Also, an IDQ detector with a low dead time $\tau \approx 15 \ \mu$s has a high afterpulse probability. It also helps verify the universality of equations.

The measurements were carried out for $\mu$ range $\mu \in [0.1, 2]$ ph/pulse with the step $\Delta \mu = 0.1$ ph/pulse. We process measured data and make a fit for the presented models according to two different $\mu$ ranges: $\mu_1 = \{0.1, \hdots 1\}$ ph/pulse and $\mu_2 \in \{0.1, \hdots 2\}$ ph/pulse. Fitted parameters for these different sets will differ, and we can observe the bounds of applicability of the proposed models. 

\begin{figure}[h]\centering
	\includegraphics[width=0.45\textwidth]{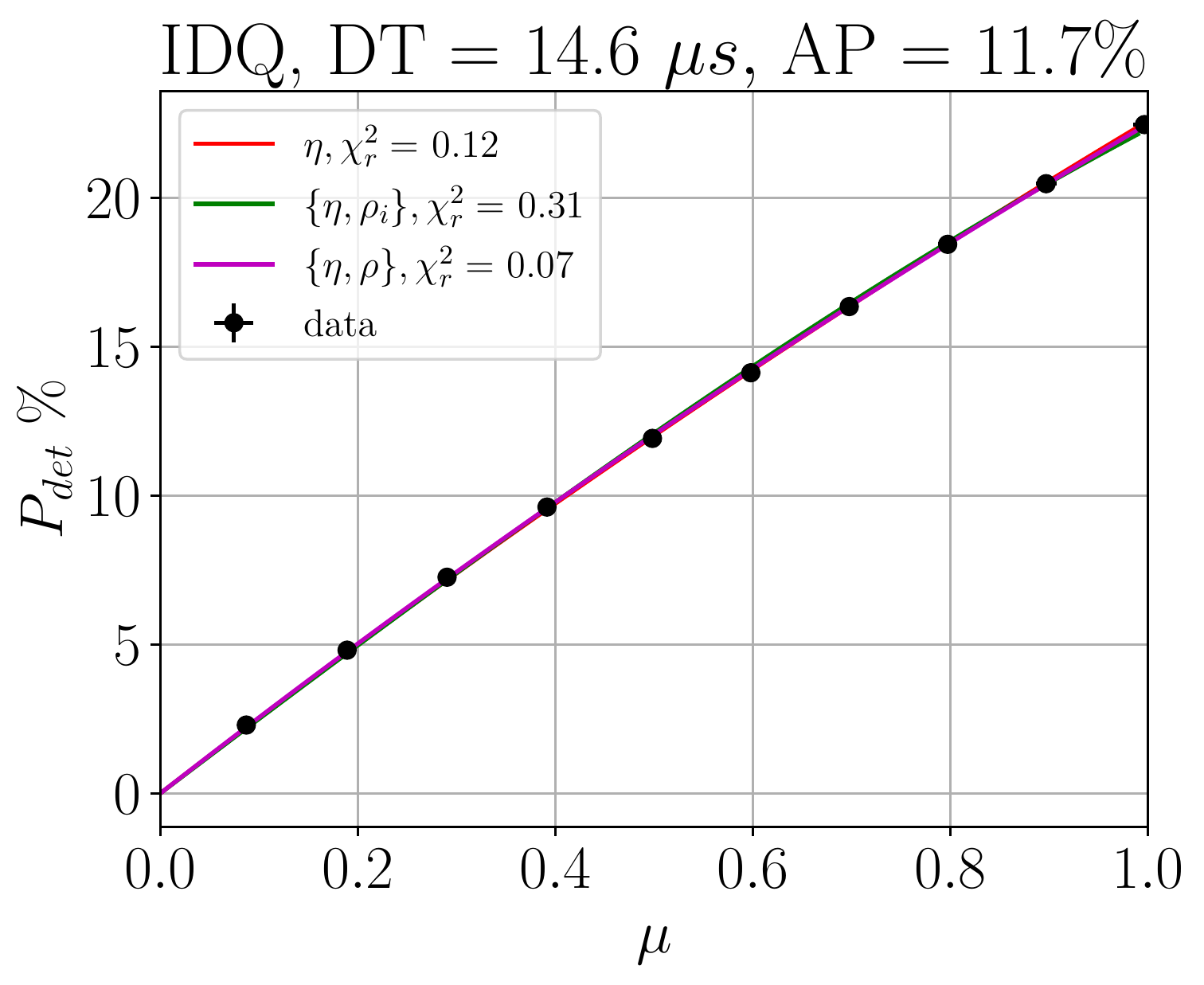} \hspace{3mm}
	\includegraphics[width=0.45\textwidth]{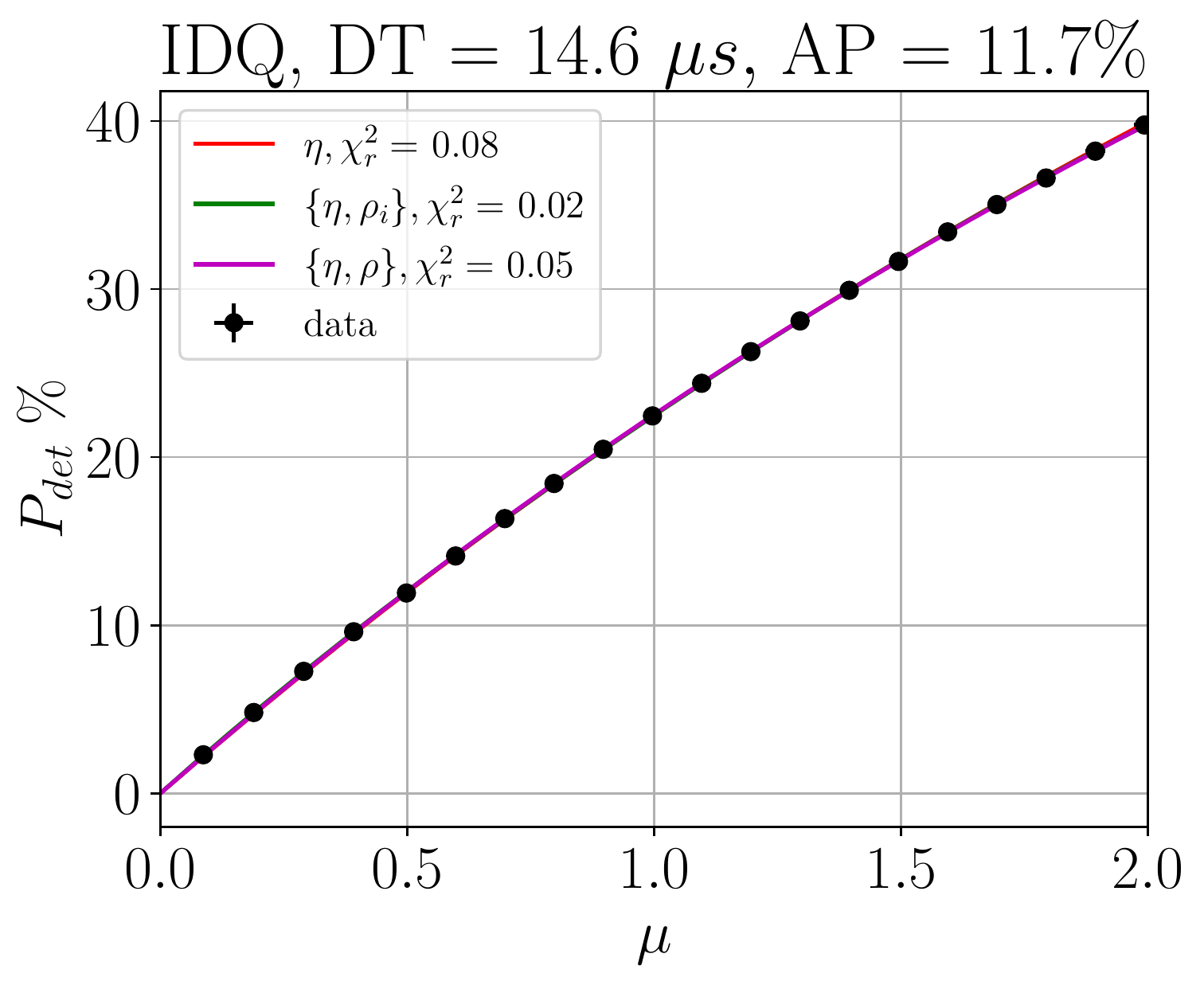}
   \caption{Fitted curves for IDQ with parameters $\tau = 14.6 \ \mu s$ and $P_{ap} = 11.7 \ \%$ for different $\mu$ ranges: a) $0 \to 1$ ph/pulse b) $0 \to 2$ ph/pulse.}
   \label{fig:fit_idq_1}
\end{figure}

\begin{figure}[h]\centering
	\includegraphics[width=0.45\textwidth]{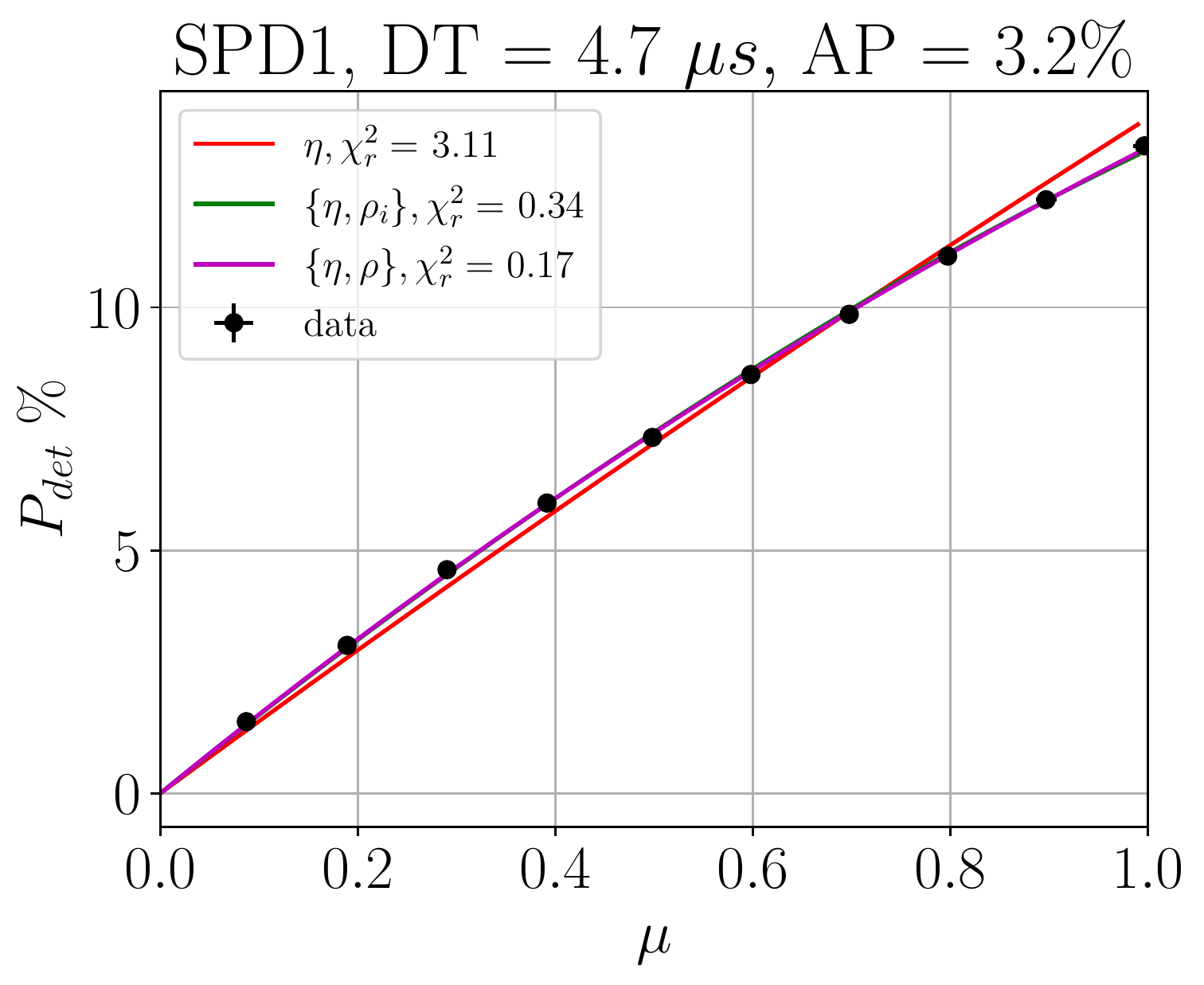} \hspace{3mm}
	\includegraphics[width=0.45\textwidth]{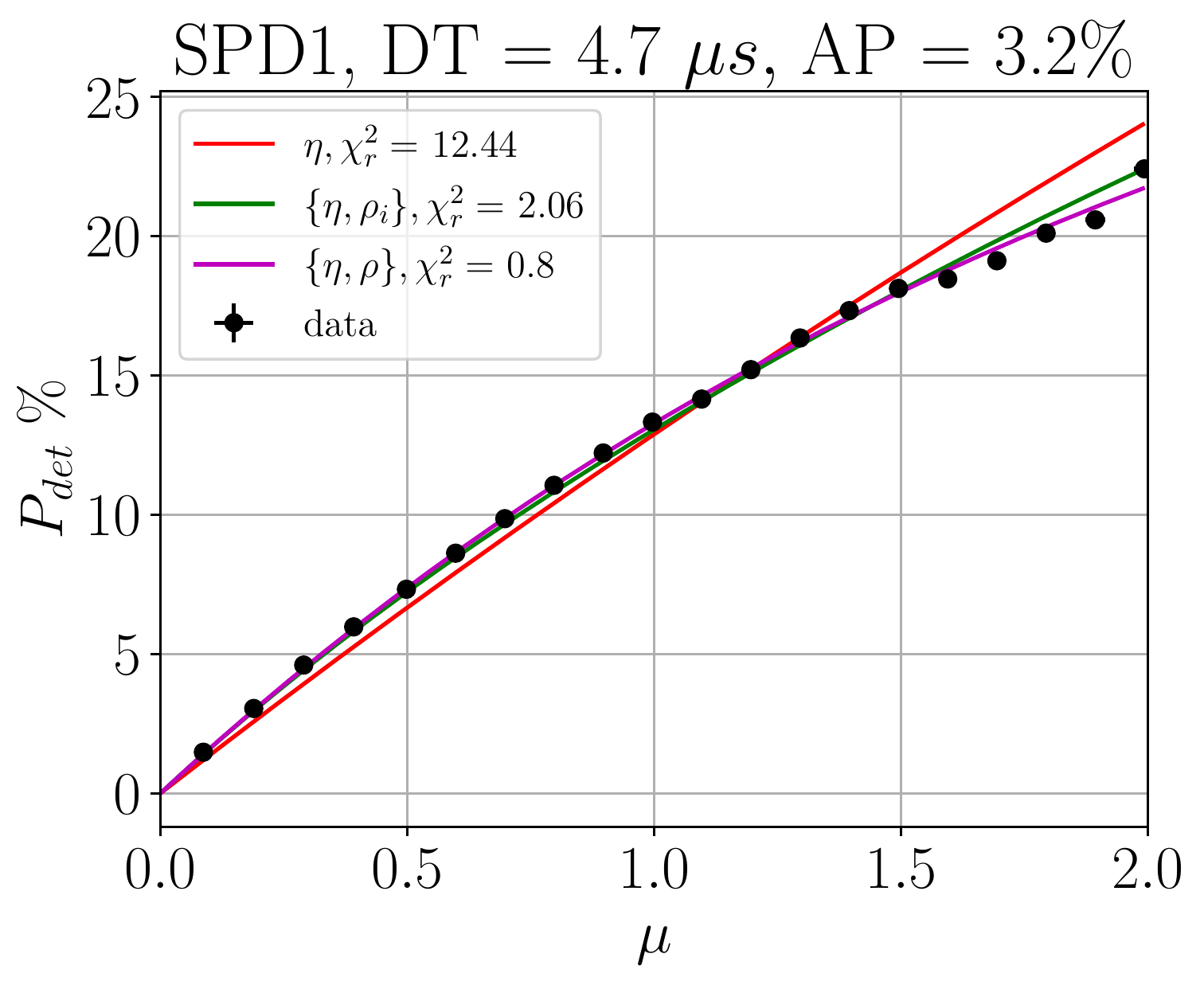}
	\caption{Fitted curves for SPD1 with parameters $\tau = 4.7 \ \mu s$ and $P_{ap} = 3.2 \ \%$ for different $\mu$ ranges: a) $0 \to 1$ ph/pulse b) $0 \to 2$ ph/pulse.}
   \label{fig:fit_spd1_1}
\end{figure}

\begin{figure}[h]\centering
	\includegraphics[width=0.45\textwidth]{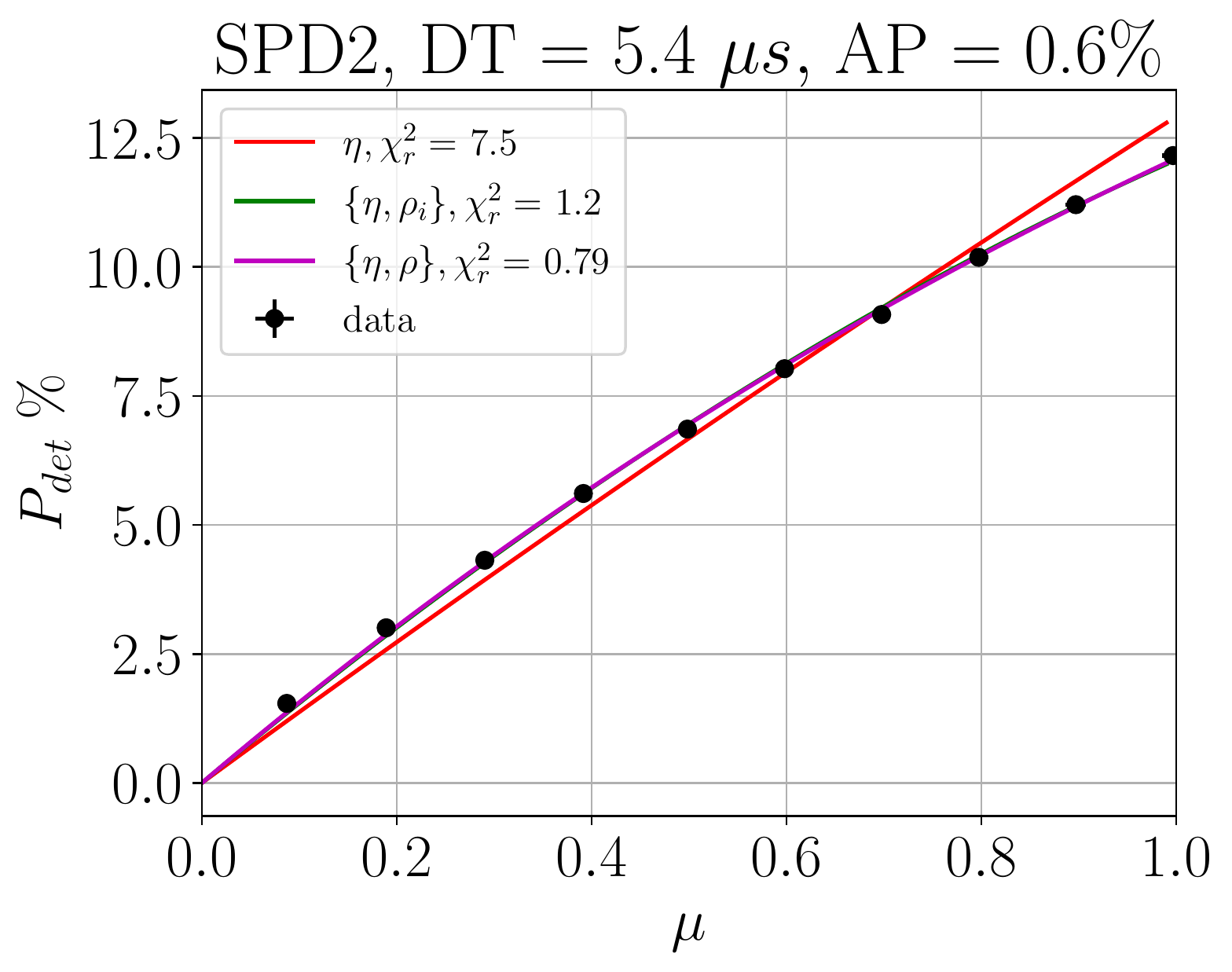} \hspace{3mm}
	\includegraphics[width=0.45\textwidth]{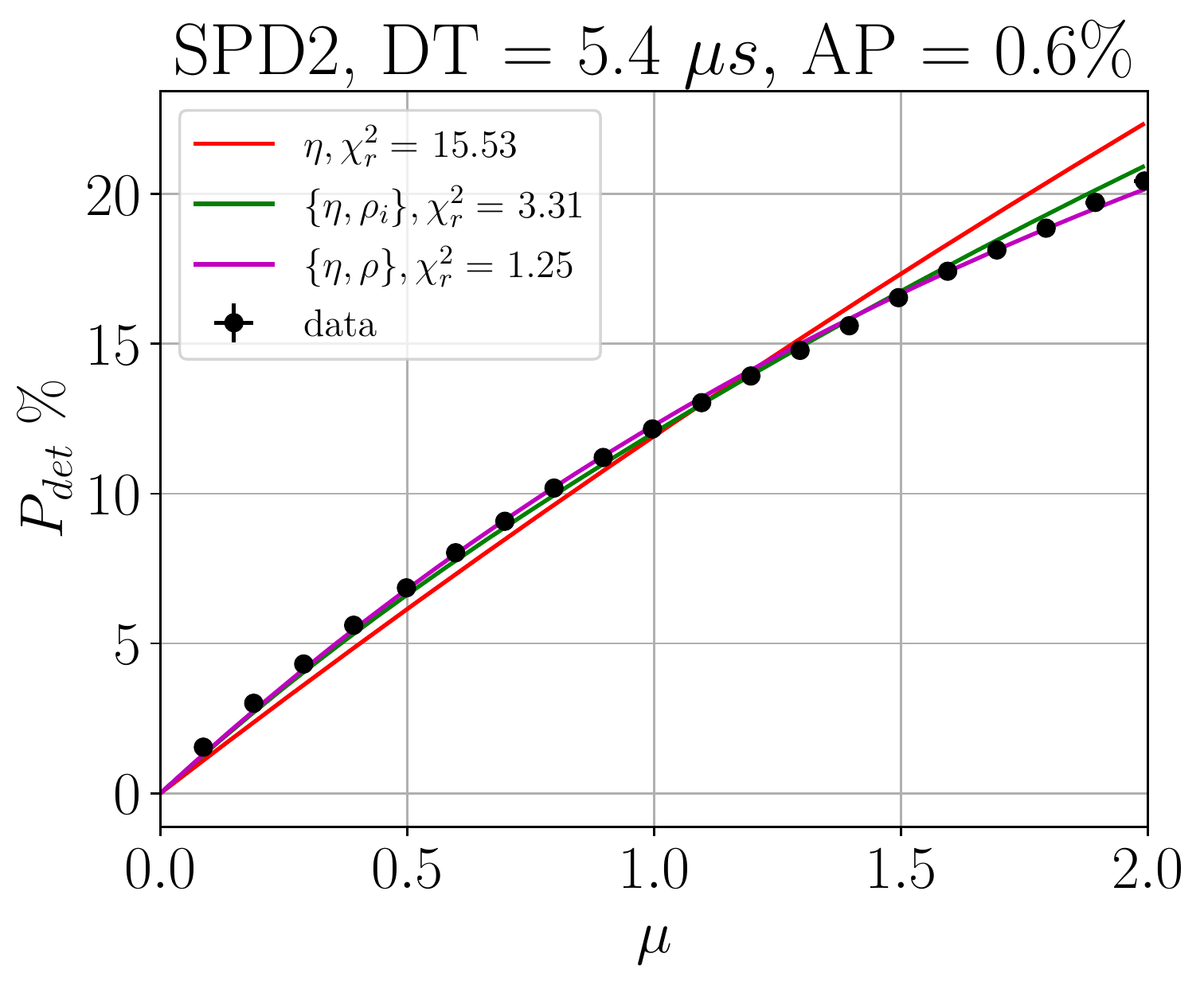}
	\caption{Fitted curves for SPD2 with parameters $\tau = 5.4 \ \mu s$ and $P_{ap} = 0.6 \ \%$ for different $\mu$ ranges: a) $0 \to 1$ ph/pulse b) $0 \to 2$ ph/pulse.}
   \label{fig:fit_spd2_1}
\end{figure}

On the pictures \ref{fig:fit_idq_1}, \ref{fig:fit_spd1_1}, \ref{fig:fit_spd2_1} we can see the experimental data, processed according to equation \ref{eq:pdet} and three fitted curves. The label $\eta$ denotes Independent model, $\{\eta, \rho_i \}$ -- Dependent model, $\{\eta, \rho \}$ -- Empirical model. The $\chi^2_r$ denotes the reduced $\chi^2$ value for presented model.

On the figure \ref{fig:fit_idq_1} for IDQ, we can see that all described models can be used to approximate the PDE of the laser pulses because $\chi^2_r < 1$. It means that for simplicity, we can assume interaction of the photons in the IDQ diode independent.

On the figure \ref{fig:fit_spd1_1} for SPD1, we can see that in the $\mu_1$ range, Dependent and Empirical models give a good result while the Independent model is on the border of applicability. If we consider the $\mu_2$ range, then the Independent model gives a significantly different result from the experiment. However, we can use two other models. The empirical model gives the best fit result.

On the figure \ref{fig:fit_spd2_1} for SPD2 we can see, that Independent model can't be used to approximate experimental data in both ranges $\mu_1$ and $\mu_2$. Dependent model shows badly for $\mu_2$ range, but good for $\mu_1$. Empirical model approximate experimental data in ranges $\mu_1$ and $\mu_2$ equally well.

We calculated the fitted parameters for the first three and six photon detection efficiencies for all models for the ranges $\mu_1$ and $\mu_2$ and presented at the tables \ref{table:mu1} and \ref{table:mu2} (the corresponding models parameters presented at the appendix \ref{ap:fitted} in tables \ref{table:par1} and \ref{table:par2}). 

From these tables, we can see that $\eta_1$ can differ for different models up to $0.02 -- 0.03$, that converted to PDE like $\Delta \eta = 2 - 3 \ \%$, which is a big enough value. Thus, for an accurate description of the detector's parameters, it is necessary to indicate within which model its PDE was determined.

Based on the obtained results, we can make several statements:

\begin{itemize}
	\item The independent model approximate experimental data bad enough. It means that there are collective photon effects inside the SPAD structure; 
	\item Using the Dependent model is more physically  grounded, but many experiments needed to obtain all required $\rho_i$ parameters. If the interesting $\mu$ range is $[0.1 ,1]$ ph/pulse, that application of this model is recommended. If $\mu > 1$ ph/pulse, then use of this model is impractical;
	\item Using the Empirical model can be convenient for large $\mu$ ranges like $\mu_2$ because it required only two empirical parameters and  approximate experimental data quite well.
\end{itemize}

\begin{figure}[h]\centering
	\includegraphics[width=0.45\textwidth]{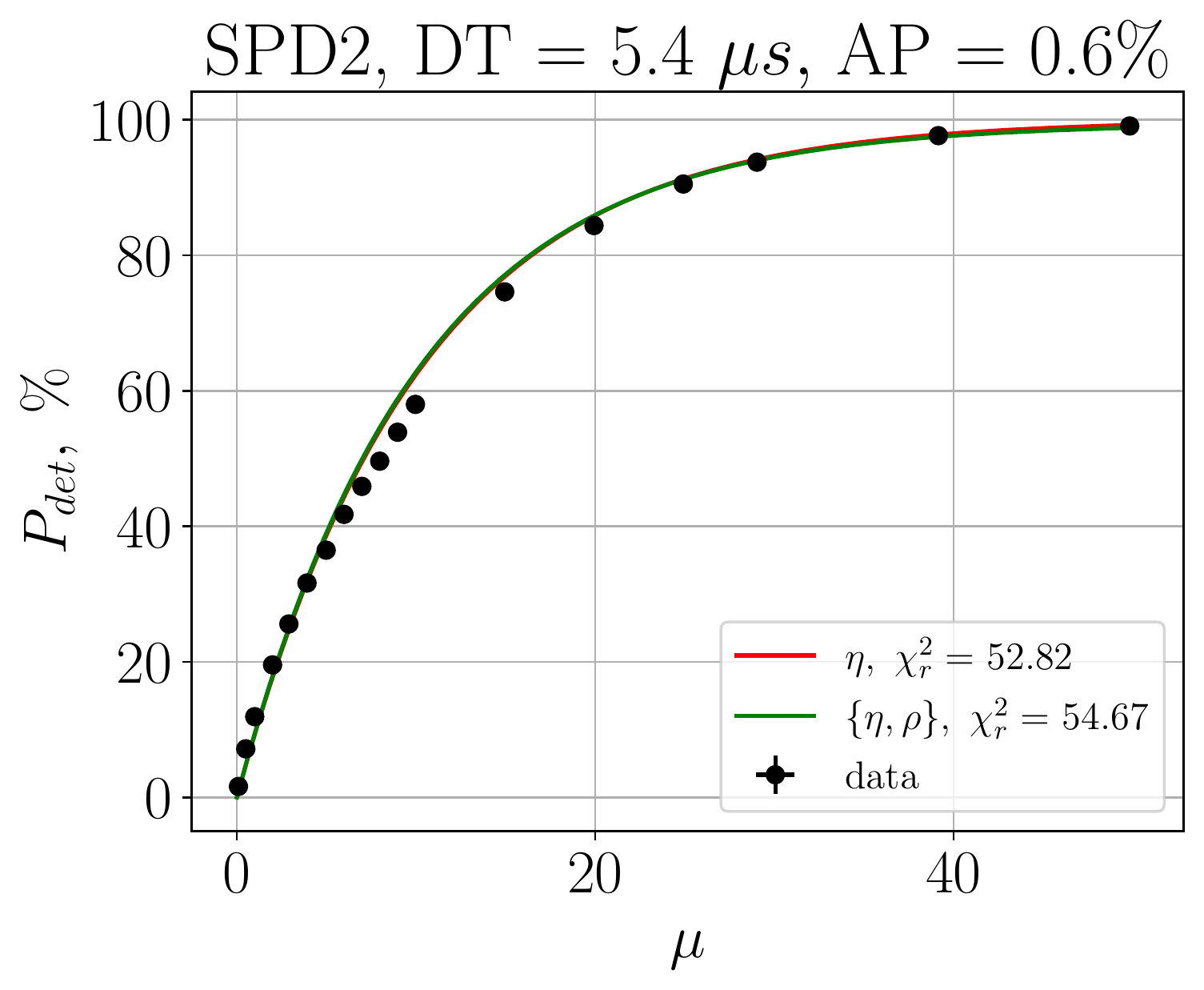} \hspace{3mm}
	\includegraphics[width=0.45\textwidth]{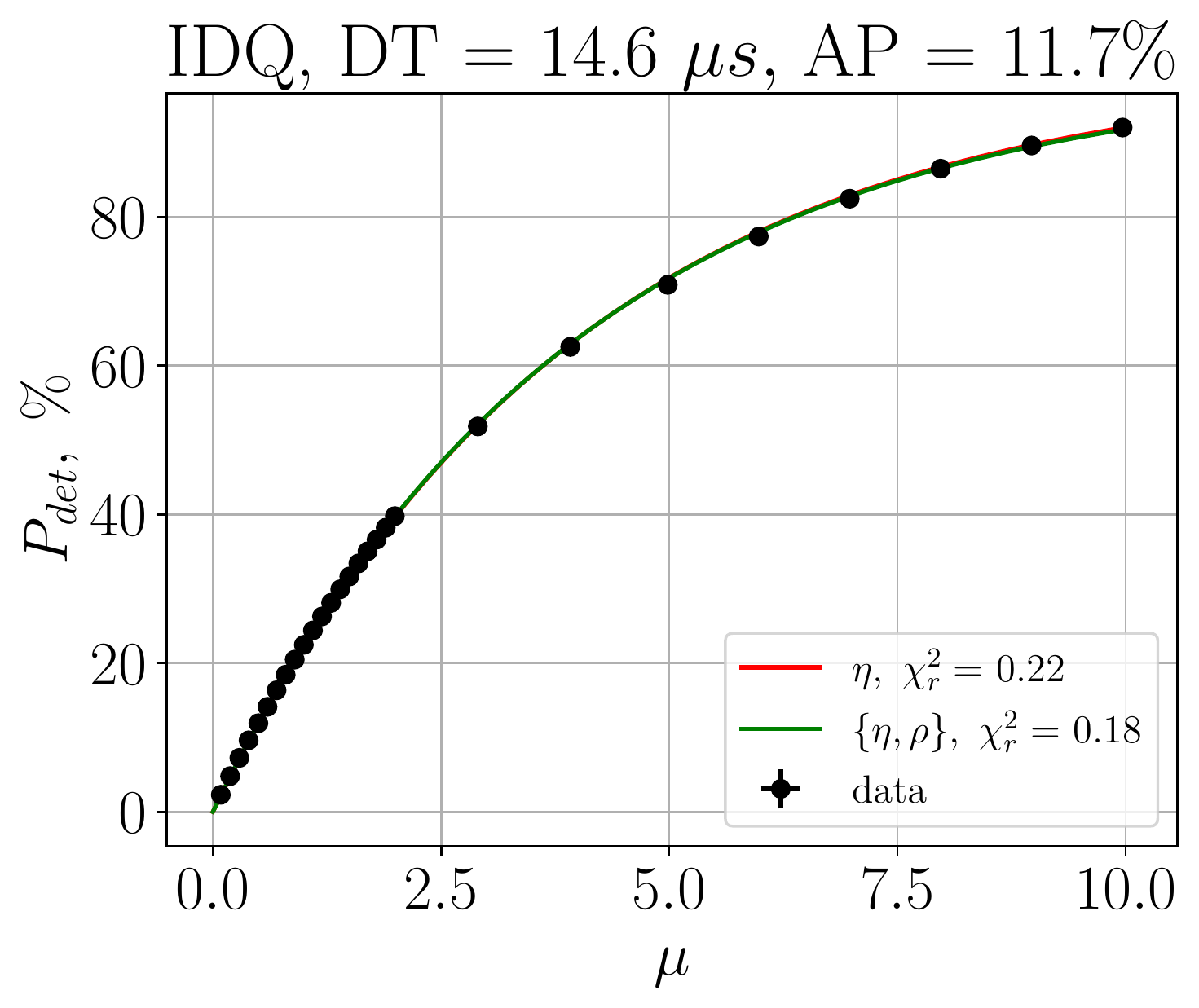}
	\caption{Compare the first and second approach to afterpulse approximation for large range of $\mu$ for: a) SPD2 with parameters $\tau = 5.4 \ \mu s$ and $P_{ap} = 0.6 \ \%$; b) IDQ with parameters $\tau = 14.6 \ \mu s$ and $P_{ap} = 11.7 \ \%$.}
   \label{fig:fit_large}
\end{figure}

Also we try to apply the Independent and Empirical models to approximate for the large $\mu$ range: $\mu = \{0.1, \hdots 50 \}$ for SPD2 and $\mu = \{0.1, \hdots 10 \}$ ph/pulse for IDQ. The obtained results presented on the picture \ref{fig:fit_large}.

From the figure \ref{fig:fit_large} we can conclude that usage of Independent or Empirical model allows quite rough approximate the experimental data, and the obtained parameters for $\eta$ and $\{\eta, \rho \}$ will sufficiently differ from the small ranges, like $\mu_1$ and $\mu_2$. However, perhaps for some specific tasks, this approach will be quite sufficient.

\begin{table}[h]
	\caption{The fitting results for range $\mu_1 = \{0.1, \hdots 1\}$ ph/pulse. }
	\label{table:mu1}
\begin{center}
	\begin{tabular}[c]{|l|l|l|l|l|l|}
		\hline
		SPD & model & $\eta_1$ & $\eta_2$ & $\eta_3$ & $\chi^2_r$ \\
		\hline
		\hline
		\multirow{3}*{IDQ} & $\eta$  & $0.256$ & $0.446$ & $0.588$ & $0.12$ \\
		\cline{2-6}
		{} & $\{\eta, \rho_i \}$  & $0.251$ & $0.469$ & $0.721$ & $0.31$ \\
		\cline{2-6}
		{} & $\{\eta, \rho \}$ & $0.259$ & $0.443$ & $0.573$ & $0.07$ \\
		\hline
		\hline
		\multirow{3}*{SPD1} & $\eta$  & $0.149$ & $0.276$ & $0.384$ & $3.1$ \\
		\cline{2-6}
		{} & $\{\eta, \rho_i \}$ & $0.164$ & $0.250$ & $0.415$ & $0.34$ \\
		\cline{2-6}
		{} & $\{\eta, \rho \}$  & $0.166$ & $0.254$ & $0.302$ & $0.17$ \\
		\hline
		\hline
		\multirow{3}*{SPD2} & $\eta$  & $0.138$ & $0.257$ & $0.360$ & $7.5$ \\
		\cline{2-6}
		{} & $\{\eta, \rho_i \}$ &  $0.159$ & $0.225$ & $0.338$ & $1.2$ \\
		\cline{2-6}
		{} & $\{\eta, \rho \}$ &  $0.161$ & $0.225$ & $0.251$ & $0.79$ \\
		\hline
	\end{tabular}
\end{center}
\end{table}

\begin{table}[h]
	\caption{The fitting results for range $\mu_2 = \{0.1, \hdots 2\}$ ph/pulse. }
	\label{table:mu2}
\begin{center}
	\begin{tabular}[c]{|l|l|l|l|l|l|l|l|l|}
		\hline
		SPD & model & $\eta_1$ & $\eta_2$ & $\eta_3$ &  $\eta_4$ & $\eta_5$ & $\eta_6$ & $\chi^2_r$ \\
		\hline
		\hline
		\multirow{3}*{IDQ} & $\eta$  & $0.255$ & $0.445$ & $0.586$ & $0.692$ & $0.770$ & $0.829$ & $0.08$ \\
		\cline{2-9}
		{} & $\{\eta, \rho_i \}$  & $0.264$ & $0.406$ & $0.644$ & $0.691$ & $0.770$ & $0.834$ & $0.02$ \\
		\cline{2-9}
		{} & $\{\eta, \rho \}$  & $0.257$ & $0.445$ & $0.583$ & $0.685$ & $0.759$ & $0.814$ & $0.05$ \\
		\hline
		\hline
		\multirow{3}*{SPD1} & $\eta$  & $0.138$ & $0.256$ & $0.358$ & $0.447$ & $0.523$ & $0.588$ & $12.44$ \\
		\cline{2-9}
		{} & $\{\eta, \rho_i \}$  & $0.164$ & $0.233$ & $0.307$ & $0.402$ & $0.495$ & $0.577$ & $2.06$ \\
		\cline{2-9}
		{} & $\{\eta, \rho \}$ & $0.164$ & $0.255$ & $0.305$ & $0.333$ & $0.349$ & $0.357$ & $0.8$ \\
		\hline
		\hline
		\multirow{3}*{SPD2} & $\eta$  & $0.127$ & $0.237$ & $0.334$ & $0.419$ & $0.492$ & $0.557$ & $15.53$ \\
		\cline{2-9}
		{} & $\{\eta, \rho_i \}$  & $0.150$ & $0.214$ & $0.291$ & $0.384$ & $0.473$ & $0.552$ & $3.31$ \\
		\cline{2-9}
		{} & $\{\eta, \rho \}$ & $0.152$ & $0.237$ & $0.284$ & $0.311$ & $0.326$ & $0.334$ & $1.25$ \\
		\hline
	\end{tabular}
\end{center}
\end{table}

\newpage

\section{Conclusion}

\qquad The present work investigates whether the detection of $n$ photons on the SPAD can be considered independent events. We conclude that in the general case, no. Even on small ranges $\mu \in [0, 1]$, for example, in the case of the SPD2 detector, the model poorly approximates the experimental data. However, this model finds strong confirmation on the IDQ detector operating in freerunning mode. The Independent model's available range of use is $\mu \in [0, 10]$. Perhaps, in the gated mode, detectors arise some physical processes that affect the probability of detecting $n$ photons states. This statement will be studied in the following works using physical SPAD models.

A dependent model was proposed, which is generally recommended to be used on the range $\mu \in [0, 1]$. To do this, we need to calculate only three empirical parameters. This model for large $\mu$ is impractical since the number of empirical parameters becomes excessively large. This model tries to describe the correlation processes in detecting $n$-photon states, which determines its value in the accurate modeling of processes in SPAD.

The empirical model, although it does not carry a physical meaning, nevertheless allows one to reasonably accurately approximate the experimental data on a sufficiently large range $\mu \in [0, 2]$ (and more). In the Dependent model on large $\mu$, problems begin with determining many empirical parameters. We recommend using this model if number of the experimental points is small.

The proposed Dependent and Empirical Models are more accurate than the classic Independent Model. Therefore, when determining a QKD system's security, where laser pulses' energy can reach $1$ ph/pulse, one of the two recommended models should be used in the general case. The use of the Independent model is acceptable if experiments have been carried out. We have proved that this model approximates the experimental data well.

\newpage

\appendix

\section{Laser pulse detection probability}\label{ap:det}

\qquad We should use the next afterpulse models to determine its influence on the total probability of the detector's trigger:

\begin{equation}\label{eq:2}
	\begin{split}
		\text{1st order}:& \ P = P_0 \frac{1}{1 - p_{ap}}, \\
		\text{2nd order}:& \ P = P_0 \frac{1}{1 - p_{ap}} - P^2_0 \frac{p_{ap}}{(1 - p_{ap})^2 (1 + p_{ap})}. \\
	\end{split}
\end{equation}

We introduce the following definitions: $\nu_{\tau} = \frac{1}{\tau}$ -- the limit count of detector's triggers (determined by the dead time $\tau$), $\nu_l$ -- the repetition rate of the laser pulses. Also, we define the $R$ -- experimentally measured triggers of the detector with laser pulses, $R^\prime$ -- experimentally measured triggers of the detectors without laser pulses.   $R^0_{sig}$ -- the number of laser pulses that can be detected, taking into account possible triggers blocked by dead time, $R_{sig}$ -- the number of laser pulses that were detected in the experiment, $R^\prime_{sig}$ -- the number of laser pulses, that were blocked by the dead time.  $R^0_{dc}$ and $R^0_{dc,d}$ is the number of dark counts that can be detected, taking into account possible triggers blocked by dead time for light on and off,  $R_{dc}$ and $R_{dc,d}$ is the thermal dark counts, measured in the experiment, for light on and off, $R^\prime_{dc}$ and $R^\prime_{dc,d}$ is the number of dark counts, that were blocked by the dead time for light on and off.

\begin{equation}\label{eq:3}
	\begin{split}
		&R^0_{sig} = R_{sig} + R^\prime_{sig},\\
		&R^0_{dc} = R_{dc} + R^\prime_{dc},\\
		&R^0_{dc,d} = R_{dc,d} + R^\prime_{dc,d}.\\
	\end{split}
\end{equation}

We should notice that $R^0_{dc, d} \approx R_{dc, d}$, because the DCR is low, and dead time blocking effect is insufficient. Also, the total DCR for light on and off are equal: $R^0_{dc} = R^0_{dc,d}$. So, we can rewrite this equations to the next:

\begin{equation}\label{eq:4}
	\begin{split}
		&R^0_{sig} = R_{sig} + R^\prime_{sig},\\
		&R_{dc,d} = R_{dc} + R^\prime_{dc},\\
	\end{split}
\end{equation}

Parameter $R^\prime_{sig}$ and $R^\prime_{dc}$ can be defined as:

\begin{equation}\label{eq:5}
	\begin{split}
		&R^\prime_{sig} = P^0_{sig} \biggl(R_{sig} \biggl[\frac{\tau}{T}\biggr] + (R - R_{sig}) \frac{\tau}{T}\biggr), \\
		&R^\prime_{dc} = P^0_{dc} R  \frac{\tau}{T}
	\end{split}
\end{equation}

with $P^0_{sig} = \frac{R^0_{sig}}{\nu_l}$ and $P^0_{dc} = P_{dc,d} = \frac{R_{dc,d}}{\nu_{\tau}}$ -- the probability of the detection of single laser pulse and probability of the occurrence of thermal noise click in time window $\tau$, $T$ -- the repetition period of laser pulses.

The integer part of the expression  $\biggl[\frac{\tau}{T}\biggr]$ in the equation for $R_{sig}$ determines the count of blocked by dead time laser pulses. If $\tau < T$, there are no laser pulses blocked by the previous laser pulses. However, if $\tau > T$, one laser pulse can block one or even more laser pulses, the noise counts are randomly distributed, and we assume that the average count of blocked laser pulses determined by the terms with expression $\frac{\tau}{T}$.

The probability of click in the time window $\tau$  excluding afterpulse effect is named $P_0$ and $P_{0 d}$ for light on and off can be determined as follows:

\begin{equation}\label{eq:6}
	\begin{split}
		&P_0 = P_{sig} + P_{dc} - P_{sig} P_{dc}, \\
		&P^\prime_{0} = P_{dc,d},
	\end{split}
\end{equation}

with $P_{sig} = \frac{R_{sig}}{\nu_{\tau}}$ -- the probability of trigger due to $R_{sig}$, $P_{dc} = \frac{R_{dc}}{\nu_{\tau}}$ -- due to $R_{dc}$,  $P_{dc,d} = \frac{R_{dc,d}}{\nu_{\tau}}$ due to $R_{dc,d}$.

\subsection{1st order model}

\qquad On the one hand, we can determine the $P$ and $P^\prime$ from $P_0$ and $P^\prime_0$, taking into account the first order of the afterpulse model from equation \ref{eq:2}:

\begin{equation}\label{eq:7}
	\begin{split}
		&P = P_0 \frac{1}{1 - p_{ap}}, \\
		&P^\prime = P^\prime_0 \frac{1}{1 - p_{ap}}.
	\end{split}
\end{equation}

On the other hand, we can derive $P$ and $P^\prime$ from statistics for light on and off:

\begin{equation}\label{eq:8}
	\begin{split}
		&P = \frac{R}{\nu_{\tau}},\\
		&P^\prime = \frac{R^\prime}{\nu_{\tau}}.
	\end{split}
\end{equation}

Transform  equations \ref{eq:7} and \ref{eq:8} using the equation \ref{eq:6}:

\begin{equation}\label{eq:9}
	\begin{split}
		&P_0 = P_{sig} + P_{dc} - P_{sig}P_{dc} = (1 - p_{ap}) \frac{R}{\nu_{\tau}}, \\
		&P^\prime_0 = P_{dc,d} = (1 - p_{ap}) \frac{R^\prime}{\nu_{\tau}}.
	\end{split}
\end{equation}

Derive the $P_{dc}$, using the equations \ref{eq:4}, \ref{eq:5}, \ref{eq:6}, \ref{eq:9}:
\begin{multline}\label{eq:10}
	P_{dc} = \frac{R_{dc}}{\nu_{\tau}} = \frac{R_{dc,d} - R^\prime_{dc}}{\nu_{\tau}} = P_{dc,d} - \frac{P^0_{dc} R \frac{\tau}{T}}{\nu_{\tau}} = P_{dc,d} - P_{dc,d} \frac{R}{\nu_{\tau}} \frac{\tau}{T} = \\
	= P^\prime_0 (1 - \frac{R}{\nu_{\tau}} \frac{\tau}{T}) = (1 - p_{ap}) \frac{R^\prime}{\nu_{\tau}}(1 - \frac{R}{\nu_{\tau}} \frac{\tau}{T})
\end{multline}

Derive the $P_{sig}$ from equation \ref{eq:9}:
\begin{equation}\label{eq:11}
	P_0 = P_{sig}(1 - P_{dc}) + P_{dc} = (1 - p_{ap}) \frac{R}{\nu_{\tau}}
\end{equation}

And using the equation \ref{eq:10}:

\begin{equation}\label{eq:12}
	P_{sig} = \frac{(1 - p_{ap})\frac{R}{\nu_{\tau}} - P_{dc} }{1 - P_{dc}} = \frac{(1 - p_{ap}) \frac{R}{\nu_{\tau}} - 1}{1 - P_{dc}} + 1 =  1 - \frac{1 - (1 - p_{ap}) \frac{R}{\nu_{\tau}}}{1 - (1 - p_{ap}) \frac{R^\prime}{\nu_{\tau}} (1 - \frac{R}{\nu_{\tau}} \frac{\tau}{T})}
\end{equation}

From $P_{sig}$ we can derive the $R_{sig} = P_{sig} \nu_{\tau}$, and as a result $R^0_{sig}$ and $P^0_{sig} = \frac{R^0_{sig}}{\nu_l}$, also using equations \ref{eq:4} and \ref{eq:5}:

\begin{equation}\label{eq:13}
	P^0_{sig} = \frac{R_{sig} + R^\prime_{sig}}{\nu_l} = \frac{R_{sig} + P^0_{sig}(R_{sig} [\frac{\tau}{T}] + (R - R_{sig}) \frac{\tau}{T})}{\nu_l} 
\end{equation}

\begin{equation}\label{eq:14}
	P^0_{sig}(\nu_l - R_{sig} [\frac{\tau}{T}] - (R - R_{sig}) \frac{\tau}{T}) = R_{sig}
\end{equation}

And the final equation can be derived as:

\begin{equation}\label{eq:15}
	\begin{split}
		&P^0_{sig} = \frac{R_{sig}}{\nu_l - R \frac{\tau}{T} + R_{sig}(\frac{\tau}{T} - [\frac{\tau}{T}])}, \\
		\text{with } &R_{sig} = \nu_{\tau} - \frac{\nu_{\tau} - (1 - p_{ap}) R}{1 - (1 - p_{ap}) \frac{R^\prime}{\nu_{\tau}} (1 - \frac{R}{\nu_{\tau}} \frac{\tau}{T})}
	\end{split}
\end{equation}

\subsection{2nd order model}

\qquad Now we will determine the $P$ and $P^\prime$ from $P_0$ and $P^\prime_0$, taking into account the second order of afterpulse model from equation \ref{eq:2}:

\begin{equation}\label{eq:16}
	\begin{split}
		&P = P_0 \frac{1}{1 - p_{ap}} - P^2_0 \frac{p_{ap}}{(1 - p_{ap})^2(1 + p_{ap})}, \\
		&P^\prime = P^\prime_0 \frac{1}{1 - p_{ap}} - P^{\prime 2}_0 \frac{p_{ap}}{(1 - p_{ap})^2(1 + p_{ap})}.
	\end{split}
\end{equation}

The evaluation of $P$ and $P^\prime$ from statistics are similar to the first-order case.  

Now the problem is how to extract the $P_0$ and $P^\prime_0$ terms from these equations. We need to solve the quadratic equation relate to $P_0$:

\begin{equation}\label{eq:17}
	P^2_0 \frac{p_{ap}}{(1 - p_{ap})^2(1 + p_{ap})} - P_0 \frac{1}{1 - p_{ap}} + P = 0
\end{equation}

Calculate the discriminant:

\begin{equation}\label{eq:18}
	D = \frac{1}{(1 - p_{ap})^2} - \frac{4 p_{ap} P}{(1 - p_{ap})^2 (1 + p_{ap})} = \frac{1}{(1 - p_{ap})^2} \biggl(1 - \frac{4 p_{ap} P}{1 + p_{ap}}\biggr)
\end{equation}

\begin{equation}\label{eq:19}
	P^{\pm}_0 = \frac{(1 - p_{ap})^2 (1 + p_{ap}) (\frac{1}{1 - p_{ap}} \pm \frac{1}{1 - p_{ap}} \sqrt{1 - \frac{4 p_{ap} P}{1 + p_{ap}}}) }{2 p_{ap}}
\end{equation}

\begin{equation}\label{eq:20}
	P^{\pm}_0 = \frac{1 - p^2_{ap}}{2p_{ap}} \biggl(1 \pm \sqrt{1 - \frac{4 p_{ap} P}{1 + p_{ap}}}\biggr)
\end{equation}

The $P^+_0$ value is not an adequate solution. We need to use $P^-_0$ as $P_0$:

\begin{equation}\label{eq:21}
	P_0 = \frac{1 - p^2_{ap}}{2p_{ap}} \biggl(1 - \sqrt{1 - \frac{4 p_{ap} P}{1 + p_{ap}}}\biggr)
\end{equation}

The analogy equation can be derived for $P^\prime_0$ value, but we need to change $P$ to $P^\prime$.

The new expression for $P_0$ and $P^\prime_0$ will change the expressions for $P_{dc}$ and $P_{sig}$. Derive the $P_{dc}$, using the equations \ref{eq:8}, \ref{eq:10}, \ref{eq:21}:

\begin{equation}\label{eq:22}
		P_{dc} = P^\prime_0 (1 - \frac{R}{\nu_{\tau}} \frac{\tau}{T}) = \frac{1 - p^2_{ap}}{2p_{ap}} \biggl(1 - \sqrt{1 - \frac{4 p_{ap} \frac{R^\prime}{\nu_{\tau}}}{1 + p_{ap}}}\biggr) (1 - \frac{R}{\nu_{\tau}} \frac{\tau}{T})
\end{equation}

Derive the $P_{sig}$ using the equations \ref{eq:8}, \ref{eq:11}, \ref{eq:21}, \ref{eq:22}:

\begin{equation}\label{eq:23}
	P_{sig} = \frac{P_0 - P_{dc}}{1 - P_{dc}} = 1 - \frac{1 - P_0}{1 - P_{dc}} = 1 - \frac{1 - \frac{1 - p^2_{ap}}{2p_{ap}} \biggl(1 - \sqrt{1 - \frac{4 p_{ap} \frac{R}{\nu_{\tau}}}{1 + p_{ap}}}\biggr)}{1 - \frac{1 - p^2_{ap}}{2p_{ap}} \biggl(1 - \sqrt{1 - \frac{4 p_{ap} \frac{R^\prime}{\nu_{\tau}}}{1 + p_{ap}}}\biggr) (1 - \frac{R}{\nu_{\tau}} \frac{\tau}{T})}
\end{equation}

And the final equation can be derived as:

\begin{equation}\label{eq:24}
	\begin{split}
		&P^0_{sig} = \frac{R_{sig}}{\nu_l - R \frac{\tau}{T} + R_{sig}(\frac{\tau}{T} - [\frac{\tau}{T}])}, \\
		\text{with } &R_{sig} = \nu_{\tau} - \nu_{\tau} \frac{1 - \frac{1 - p^2_{ap}}{2p_{ap}} \biggl(1 - \sqrt{1 - \frac{4 p_{ap} \frac{R}{\nu_{\tau}}}{1 + p_{ap}}}\biggr)}{1 - \frac{1 - p^2_{ap}}{2p_{ap}} \biggl(1 - \sqrt{1 - \frac{4 p_{ap} \frac{R^\prime}{\nu_{\tau}}}{1 + p_{ap}}}\biggr) (1 - \frac{R}{\nu_{\tau}} \frac{\tau}{T})}
	\end{split}
\end{equation}

\section{$\chi^2$ Detailed calculation algorithm}\label{ap:chi2}

\qquad The $\chi^2$ function can be derived as follows:

\begin{equation}
	\chi^2 = \sum^N_{i=1}\frac{(R^{0}_{sig(i)} - R^{th}_i)^2}{\sigma^2_{R^0_{sig(i)}} + \sigma^2_{p_i}}.
\end{equation}

Now we will describe in detail how we can calculate each parameter. We have the $N$ experimental points. In each of that, we have different $W_i$ and $R_i$ values. To accurately determine the $R^{0}_{sig(i)}$ we need to collect statistics about $n = 300$ measurements, and process this values as:

\begin{equation}
	R_i = \frac{1}{n} \sum^n_{j=1} R_{(i)j}.
\end{equation}

After that, we can determine the standard deviation of the collected statistics:

\begin{equation}
	\sigma_{R_i} = \frac{1}{\sqrt{n}} \sqrt{\frac{\sum^n_{j=1} (R_{(i)j} - R_i )^2}{n-1}},
\end{equation}

where the first term $\frac{1}{\sqrt{n}}$ placed because we need to calculate the standard deviation of the mean value $R_i$, but not the value $R_{(i)j}$ itself.

The parameter $R^0_{sig}$ and $\sigma_{R^0_{sig(i)}}$ can be determined from the equation \ref{eq:pdet}.

The parameter $R^{th}_{i}$ denotes the theoretical value according to the model obtained at the point with energy $W_i$. The parameter $\sigma_{p_i}$ denotes the standard deviation of the theoretical model, which arose due to the mean photon's uncertainties per pulse. We can calculate it as:

\begin{equation}
	\sigma^2_{p_i} = \sigma^{2}_{R^{th}_i}(W_i) + \sigma^{2}_{R^{th}_i} (\alpha),
\end{equation}

where $\sigma^{th}_{R_i}(W_i)$ and $\sigma^{th}_{R_i} (\alpha)$ is  the standard deviation of the theoretical model value due to power  and attenuation. They can be derived as:

\begin{equation}
	\begin{split}
		\sigma_{R^{th}_i} (W_i) &= \Delta W_i \frac{\partial R^{th}_i(W)}{\partial W}, \\
		\sigma_{R^{th}_i} (\alpha) &= \Delta \alpha \frac{\partial R^{th}_i (\alpha)}{\partial \alpha},
	\end{split}
\end{equation}

where $\Delta W$ and the $\Delta \alpha$ is the standard deviation for the power meter ($\Delta W = 200$ pW) and attenuation ($\delta \alpha = 0.1$).

\ref{ap:fitted} in tables \ref{table:par1} and \ref{table:par2}). 

\newpage

\section{Fitted parameters}\label{ap:fitted}

\begin{table}[h]
	\caption{The fitted parameters for range $\mu_1 = \{0.1, \hdots 1\}$ ph/pulse. }
	\label{table:par1}
\begin{center}
	\begin{tabular}[c]{|l|l|l|l|l|l|l|}
		\hline
		SPD & model & $\eta$ & $\rho$ & $\rho_2$ & $\eta_3$ & $\chi^2_r$ \\
		\hline
		\hline
		\multirow{3}*{IDQ} & $\eta$  & $0.256$ & $-$ & $-$ & $-$ & $0.12$ \\
		\cline{2-7}
		{} & $\{\eta, \rho_i \}$  & $0.251$ & $-$ & $1.157$ & $1.884$ & $0.31$ \\
		\cline{2-7}
		{} & $\{\eta, \rho \}$ & $0.259$ & $1.117$ & $-$ & $-$ & $0.07$ \\
		\hline
		\hline
		\multirow{3}*{SPD1} & $\eta$  & $0.149$ & $-$ & $-$ & $-$ & $3.1$ \\
		\cline{2-7}
		{} & $\{\eta, \rho_i \}$ & $0.164$ & $-$ & $0.624$ & $1.334$ & $0.34$ \\
		\cline{2-7}
		{} & $\{\eta, \rho \}$  & $0.166$ & $2.81$ & $-$ & $-$ & $0.17$ \\
		\hline
		\hline
		\multirow{3}*{SPD2} & $\eta$  & $0.138$ & $-$ & $-$ & $-$ & $7.5$ \\
		\cline{2-7}
		{} & $\{\eta, \rho_i \}$ &  $0.159$ & $-$ & $0.5$ & $0.917$ & $1.2$ \\
		\cline{2-7}
		{} & $\{\eta, \rho \}$ &  $0.161$ & $3.726$ & $-$ & $-$ & $0.79$ \\
		\hline
	\end{tabular}
\end{center}
\end{table}

\begin{table}[h]
	\caption{The fitted parameters for range $\mu_2 = \{0.1, \hdots 2\}$ ph/pulse. }
	\label{table:par2}
\begin{center}
	\begin{tabular}[c]{|l|l|l|l|l|l|l|l|l|l|}
		\hline
		SPD & model & $\eta$ & $\rho$ & $\rho_2$ & $\rho_3$ &  $\rho_4$ & $\rho_5$ & $\rho_6$ & $\chi^2_r$ \\
		\hline
		\hline
		\multirow{3}*{IDQ} & $\eta$  & $0.255$ & $-$ & $-$ & $-$ & $-$ & $-$ & $-$ & $0.08$ \\
		\cline{2-10}
		{} & $\{\eta, \rho_i \}$  & $0.264$ & $-$ & $0.731$ & $1.516$ & $0.5$ & $0.959$ & $1.061$ & $0.02$ \\
		\cline{2-10}
		{} & $\{\eta, \rho \}$  & $0.257$ & $1.038$ & $-$ & $-$ & $-$ & $-$ & $-$ & $0.05$ \\
		\hline
		\hline
		\multirow{3}*{SPD1} & $\eta$  & $0.138$ & $-$ & $-$ & $-$ & $-$ & $-$ & $-$  & $12.44$ \\
		\cline{2-10}
		{} & $\{\eta, \rho_i \}$  & $0.164$ & $-$ & $0.5$ & $0.589$ & $0.828$ & $0.949$ & $0.989$ & $2.06$ \\
		\cline{2-10}
		{} & $\{\eta, \rho \}$ & $0.164$ & $2.717$ & $-$ & $-$ & $-$ & $-$ & $-$ & $-$ \\
		\hline
		\hline
		\multirow{3}*{SPD2} & $\eta$  & $0.127$ & $-$ & $-$ & $-$ & $-$ & $-$ & $-$ & $15.53$ \\
		\cline{2-10}
		{} & $\{\eta, \rho_i \}$  & $0.150$ & $-$ & $0.5$ & $0.653$ & $0.868$ & $0.963$ & $0.993$ & $3.31$ \\
		\cline{2-10}
		{} & $\{\eta, \rho \}$ & $0.152$ & $2.897$ &  $-$ & $-$ & $-$ & $-$ & $-$ & $1.25$ \\
		\hline
	\end{tabular}
\end{center}
\end{table}

\newpage

\bibliography{draft_v1.bbl}

\end{document}